%% file: main.tex
\documentclass[]{spie}

\usepackage{amsmath,amsfonts,amssymb}
\usepackage{graphicx}
\usepackage[colorlinks=true, allcolors=blue]{hyperref}
\input{defs}

\usepackage{caption} 
\usepackage{textcomp}
\title{BICEP Array cryostat and mount design}

\author[a]{Michael Crumrine}
\author[b]{P.~A.~R.~Ade}
\author[c]{Z.~Ahmed}
\author[d]{R.~W.~Aikin}
\author[e]{K.~D.~Alexander}
\author[e]{D.~Barkats}
\author[f]{S.~J.~Benton}
\author[g]{C.~A.~Bischoff}
\author[d,h]{J.~J.~Bock}
\author[e]{R.~Bowens-Rubin}
\author[d]{J.~A.~Brevik}
\author[e]{I.~Buder}
\author[i]{E.~Bullock}
\author[e,j]{V.~Buza}
\author[e]{J.~Connors}
\author[e]{J.~Cornelison}
\author[h]{B.~P.~Crill}
\author[e]{M.~Dierickx}
\author[k]{L.~Duband}
\author[j]{C.~Dvorkin}
\author[l,m]{J.~P.~Filippini}
\author[a]{S.~Fliescher}
\author[n]{J.~Grayson}
\author[a]{G.~Hall}
\author[o]{M.~Halpern}
\author[e]{S.~Harrison}
\author[d,h]{S.~R.~Hildebrandt}
\author[p]{G.~C.~Hilton}
\author[d]{H.~Hui}
\author[c,n,p]{K.~D.~Irwin}
\author[n]{J.~Kang}
\author[e,q]{K.~S.~Karkare}
\author[n]{E.~Karpel}
\author[r]{J.~P.~Kaufman}
\author[r]{B.~G.~Keating}
\author[d]{S.~Kefeli}
\author[n]{S.~A.~Kernasovskiy}
\author[e,j]{J.~M.~Kovac}
\author[c,n]{C.~L.~Kuo}
\author[q]{N.~A.~Larsen}
\author[a]{K.~Lau}
\author[q]{E.~M.~Leitch}
\author[d]{M.~Lueker}
\author[h]{K.~G.~Megerian}
\author[d]{L.~Moncelsi}
\author[s]{T.~Namikawa}
\author[f,t]{C.~B.~Netterfield}
\author[h]{H.~T.~Nguyen}
\author[d,h]{R.~O'Brient}
\author[c,n]{R.~W.~Ogburn~IV}
\author[g]{S.~Palladino}
\author[a,i]{C.~Pryke}
\author[e]{B.~Racine}
\author[e]{S.~Richter}
\author[a]{R.~Schwarz}
\author[d]{A.~Schillaci}
\author[q,u]{C.~D.~Sheehy}
\author[d]{A.~Soliman}
\author[e]{T.~St.~Germaine}
\author[d,h]{Z.~K.~Staniszewski}
\author[d]{B.~Steinbach}
\author[b]{R.~V.~Sudiwala}
\author[d,r]{G.~P.~Teply}
\author[c,n]{K.~L.~Thompson}
\author[n]{J.~E.~Tolan}
\author[b]{C.~Tucker}
\author[h]{A.~D.~Turner}
\author[g]{C.~Umilt\`{a}}
\author[q,v]{A.~G.~Vieregg}
\author[d]{A.~Wandui}
\author[h]{A.~C.~Weber}
\author[o]{D.~V.~Wiebe}
\author[a]{J.~Willmert}
\author[e,j]{C.~L.~Wong}
\author[n,q]{W.~L.~K.~Wu}
\author[n]{H.~Yang}
\author[c,n]{K.~W.~Yoon}
\author[d]{C.~Zhang}

\affil[a]{School of Physics and Astronomy, University of Minnesota, Minneapolis, MN 55455, USA}
\affil[b]{School of Physics and Astronomy, Cardiff University, Cardiff, CF24 3AA, United Kingdom}
\affil[c]{Kavli Institute for Particle Astrophysics and Cosmology, SLAC National Accelerator Laboratory, Menlo Park, CA 94025, USA}
\affil[d]{Department of Physics, California Institute of Technology, Pasadena, CA 91125, USA}
\affil[e]{Harvard-Smithsonian Center for Astrophysics, Cambridge, MA 02138, USA}
\affil[f]{Department of Physics, University of Toronto, Toronto, Ontario, M5S 1A7, Canada}
\affil[g]{Department of Physics, University of Cincinnati, Cincinnati, OH 45221, USA}
\affil[h]{Jet Propulsion Laboratory, Pasadena, CA 91109, USA}
\affil[i]{Minnesota Institute for Astrophysics, University of Minnesota, Minneapolis, MN 55455, USA}
\affil[j]{Department of Physics, Harvard University, Cambridge, MA 02138, USA}
\affil[k]{Service des Basses Temp\'{e}ratures, Commissariat \'{a} l’Energie Atomique, 38054 Grenoble, France}
\affil[l]{Department of Physics, University of Illinois at Urbana-Champaign, Urbana, IL 61801, USA}
\affil[m]{Department of Astronomy, University of Illinois at Urbana-Champaign, Urbana, IL 61801, USA}
\affil[n]{Department of Physics, Stanford University, Stanford, CA 94305, USA}
\affil[o]{Department of Physics and Astronomy, University of British Columbia,Vancouver, British Columbia, V6T 1Z1, Canada}
\affil[p]{National Institute of Standards and Technology, Boulder, CO 80305, USA}
\affil[q]{Kavli Institute for Cosmological Physics, University of Chicago, Chicago, IL 60637, USA}
\affil[r]{Department of Physics, University of California at San Diego, La Jolla, CA 92093, USA}
\affil[s]{Leung Center for Cosmology and Particle Astrophysics, National Taiwan University, Taipei 10617, Taiwan}
\affil[t]{Canadian Institute for Advanced Research, Toronto, Ontario, M5G 1Z8, Canada}
\affil[u]{Physics Department, Brookhaven National Laboratory, Upton, NY 11973}
\affil[v]{Department of Physics, Enrico Fermi Institute, University of Chicago, Chicago, IL 60637}

\authorinfo{Further author information: (Send correspondence to M. Crumrine)\\M. Crumrine: E-mail: crumrine@umn.edu}

\begin{document} 
\maketitle

\begin{abstract}

\biceparray\ is a cosmic microwave background (CMB) polarization experiment that
will begin observing at the South Pole in early 2019. This experiment replaces
the five \bicep2 style receivers that compose the \keckarray\ with four larger
\bicep3 style receivers observing at six frequencies from 30 to 270GHz. The
95GHz and 150GHz receivers will continue to push the already deep \bk{}
CMB maps while the 30/40GHz and 220/270GHz receivers will constrain the
synchrotron and galactic dust foregrounds respectively. Here we report on the
design  and performance of the \biceparray\ instruments focusing on the mount
and cryostat systems.
\end{abstract}


\section{INTRODUCTION}

Precision measurements of the cosmic microwave background have significantly
increased our understanding of the early universe. Beginning with the discovery
of the CMB---providing the first evidence for a Big Bang origin to the
universe\cite{penzias1965,dicke1965}---experimental cosmology has become an extremely active subfield of physics.
Subsequently the development of the $\Lambda$CDM cosmological model has seen
tremendous success in its ability to accurately and self-consistently explain
and predict observational data. The $\Lambda$CDM model is, however, still
incomplete and fails to explain the homogeneity and isotropy of our observable
universe.  These issues can be resolved by an extension to the model in which
the universe undergoes a period of exponential expansion known as inflation just after the beginning\cite{guth1981}. Various models of inflation exist, many of which predict that
such an event would have generated an observable gravitational wave background.
Interaction between these inflationary gravitational waves (IGWs) and the primordial plasma
would have produced a characteristic curl-like polarization signal
colloquially referred to as ``B-modes''. Measurement of this signal is
generally characterized by the tensor to scalar ratio $r$, which can also be
used as a measure of the energy scale of inflation as well as to differentiate
inflationary models.

\begin{figure}[h]
	\center
	\includegraphics[scale=0.5]{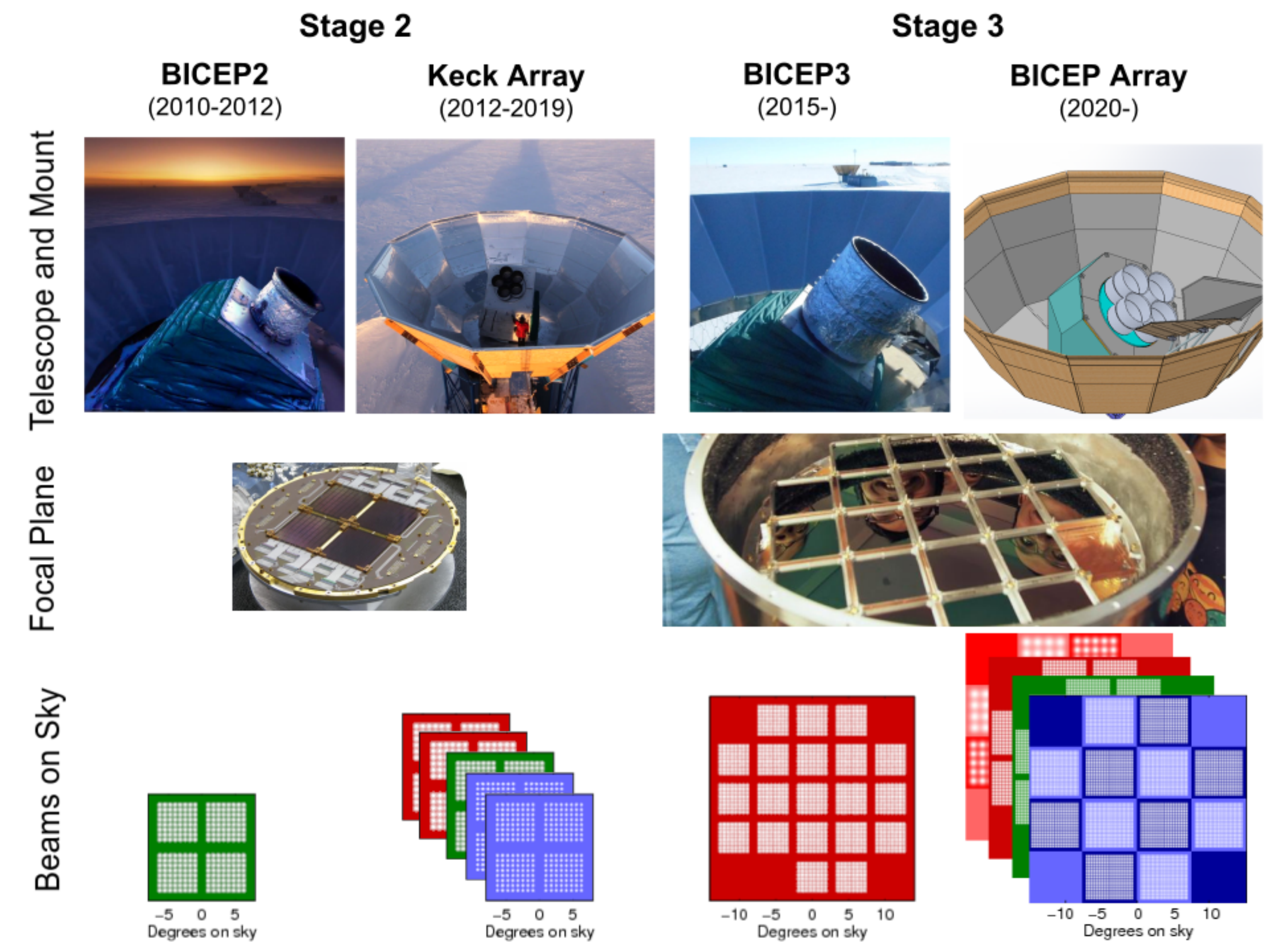}
	\caption{A graphical representation of the \bk\ experiments from
	\bicep2\ through \biceparray. The middle and bottom rows show the increased focal plane size and density of the \bicep3 style receivers as compared to \bicep2\ /\keck\ style. Different
	frequencies are represented in the bottom row by individual colors with the lowest frequency
	(30 GHz) visible in light red and the highest (270 GHz) visible in dark
	blue.}
	\label{fig:progression}
\end{figure}

Detection of the inflationary B-mode signal is complicated by the presence of
other B-mode signals. At small angular scales, lensing of divergence-like CMB
polarization can convert so called ``E-modes'' into B-modes through small
angular remappings. Other sources of B-mode signal exist in our own galaxy.
Polarized emission from galactic dust and galactic synchrotron radiation are
the two dominant foregrounds in our own galaxy, each characterized by
independent frequency and angular scale relations. Detection of any primordial
B-mode signal requires observations at multiple frequencies to enable its
separation from these foregrounds.

The \bk\ experiments are a staged series of microwave telescopes that observe
the CMB from the geographic South Pole and target degree-scale B-mode
polarization. Each successive generation of telescopes builds upon the
experience gained with the previous while simultaneously increasing
sensitivity. A graphical progression from \bicep2\ to \biceparray\ is shown
below in Fig.~\ref{fig:progression}. Foreground signals exhibit strong
frequency dependence which allows these components to be isolated from the IGW
signal by observing the microwave sky across a range of frequencies.
\keckarray\ consists of five \bicep2 style receivers\cite{ogburn2010,sheehy2010} with observations spread
across four frequency bands (95, 150, 220, and 270 GHz). The combination of
 \bicep2 and \keckarray\ data has produced the tightest constraint on
this signal to date of $r<0.07$ ($95\%$ confidence)\cite{bk_vi}. \biceparray\ will build on the success
of the \keckarray\ by deploying four \bicep3 style receivers\cite{ahmed2014}, and expanding
observations into two additional frequency bands at 30 and 40 GHz.
Extrapolating from achieved performance, \biceparray\ is projected to reach
$\sigma(r)\sim0.003$, either detecting the IGW signal or improving the constraint
to $r<0.008$ ($95\%$~confidence).

\section{Cryostat Design}

\begin{figure} [hb]
	\begin{center}
		\includegraphics[scale=0.35]{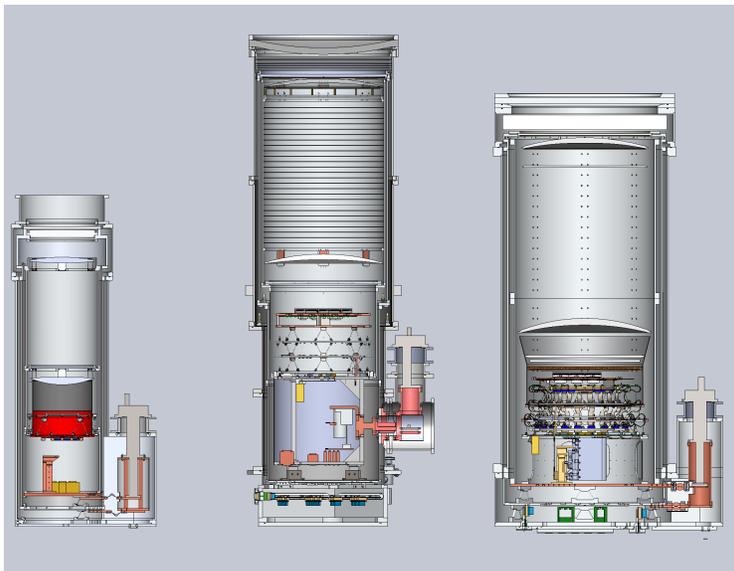}
	\end{center}
	\caption{A CAD cross section of a single \keckarray\ receiver (left), the
	\bicep3 receiver (middle), and a single
	\biceparray\ receiver (right). The \biceparray\ receivers have double
	the clear aperture and significantly increased focal plane density as
	compared to the \keckarray\ receivers. \biceparray\ also increases the
	width of the cryostat compared to \bicep3 in order to include adequate optical baffling at the
	lowest frequencies.}
	\label{fig:bavskeck}
\end{figure}

\biceparray\ continues the successful design philosophy of
the previous \bk\ receivers. A 2 meter tall vacuum shell contains two nested
stages with nominal operating temperatures of 50K and 4K which accommodate
optical elements and shield a sub-Kelvin focal plane. The design of the
sub-kelvin structure and focal plane can be found in Hui. et.
al.\cite{hui2018}, and Soliman. et. al.\cite{soliman2018}. A cross section of
the cryostat is shown in Fig.~\ref{fig:bavskeck}. The top section of the vacuum
jacket houses the vacuum window and a stack of Zotefoam filters which reduce
infrared loading onto the colder stages. More information on the performance of
this filtering scheme can be found in J.~Kang et.~al.\cite{kang2018}. The use
of low-conductivity structural supports keeps the interior stages sufficiently
supported while conducting only a small amount of heat from the room
temperature vacuum vessel.  These supports are described in more detail in
Sec.~\ref{sec:thermal_architecture}. The intermediate 50K stage serves as a
radiation shield for the interior 4K stage.  The lower $\sim70\%$ of the 50K
stage is wrapped with a $0.04''$-thick magnetic shield composed of
Amuneal\cite{amuneal} A4K\textsuperscript{\textregistered}.  The top of this
stage additionally accommodates an alumina filter to further reduce infrared
loading onto the sub-Kelvin receiver. The 4K stage serves as a second radiation
shield while also providing mounting interfaces for two alumina lenses and
optical baffling.

\begin{figure} [h]
	\begin{center}
		\includegraphics[scale=0.4]{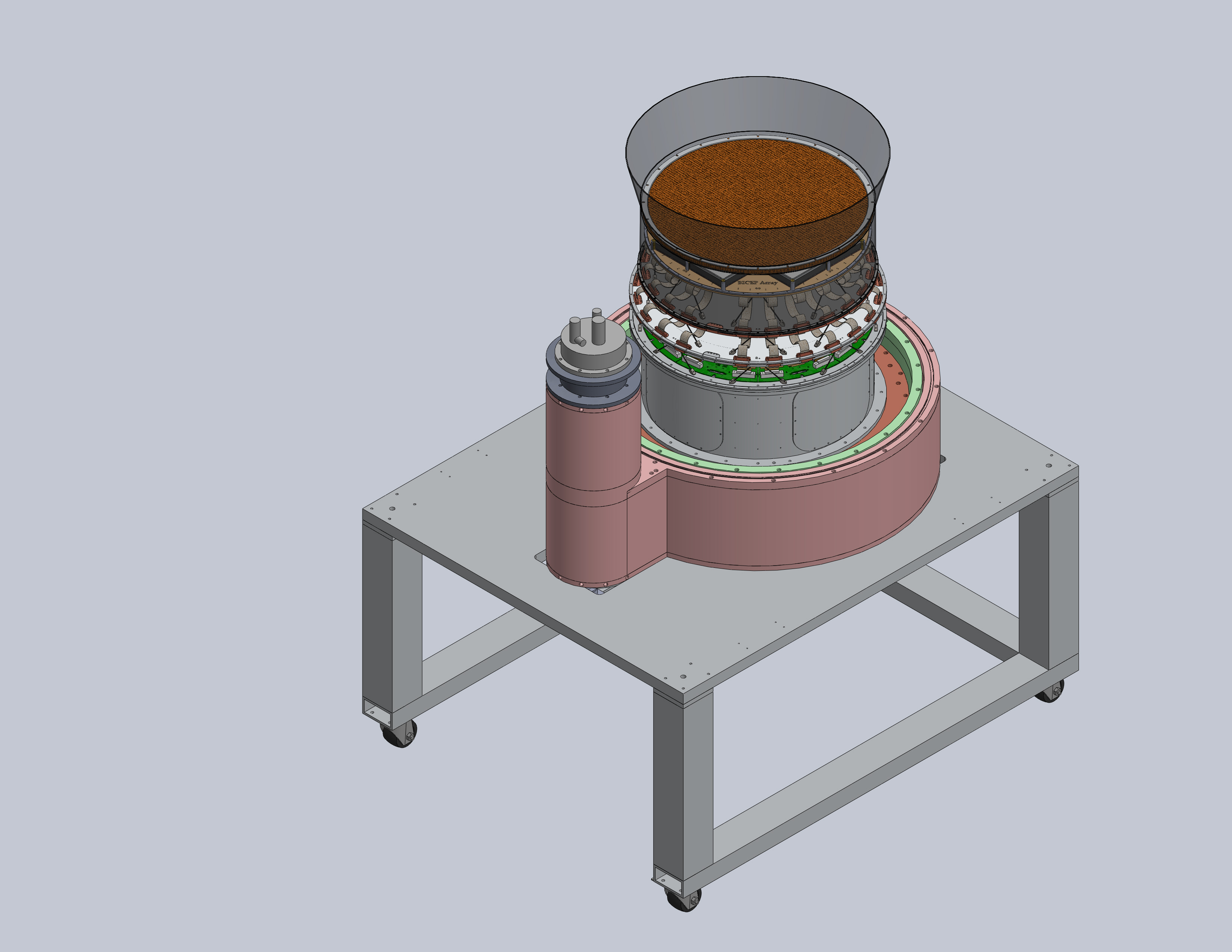}
	\end{center}
	\caption{A partially disassembled \biceparray\ cryostat. In this view
	the long upper sections of the vacuum jacket, 50K, and 4K stages have been
	removed leaving behind the base assembly that contains the pulse tube cooler,
	focal plane assembly, and the readout electronics. Elevating this base
	structure onto a lab stand allows hatches on the bottom side of the cryostat to
	be opened for access to the backend without disturbing the focal plane or any
	thermal joints.
	}
	\label{fig:base}
\end{figure}

The cryostat design enables ease of access to the focal plane assembly without
disturbing the majority of thermal joints or the back-end cabling. The
cryostat disassembles by lifting off shells successively
from the outside in, leaving behind a stand-along base which contains the
sub-Kelvin focal plane assembly, readout electronics, and the cooling system as shown in
Fig.~\ref{fig:base}. In this state, the focal plane and detector modules may be
freely accessed for maintenance. Access to the
pulse tube, heat straps, and readout cabling is provided by
hatches on the bottom side of the vacuum shell and 50K bases. This scheme
significantly reduces the time required for disassembly when accessing the
focal plane and re-assembly afterwards by allowing critical thermal junctions
and difficult part matings to remain undisrupted.

\section{Thermal Architecture}
\label{sec:thermal_architecture}

 The 50K and 4K radiation shields are cooled by the first
and second stages of a Cryomech\cite{cryomech} PT415-RM Pulse Tube cooler respectively. This
cryocooler is capable of maintaining a first stage temperature of $<45$K under
a 40W load and a second stage temperature of $<4$K under a
1.5W load. The cooler of these two stages is required to maintain a
sufficiently cold temperature to allow the operation of a three-stage helium
sorption fridge which cools the sub-Kelvin focal plane assembly. The interior
stages must therefore be well insulated in order to stay within the thermal
budget. Various aspects of the thermal architecture are described below, Table
\ref{tab:loading} provides a numerical breakdown of the estimated thermal loading onto
the cold stages.

\begin{figure}[b]
	\center
	\includegraphics{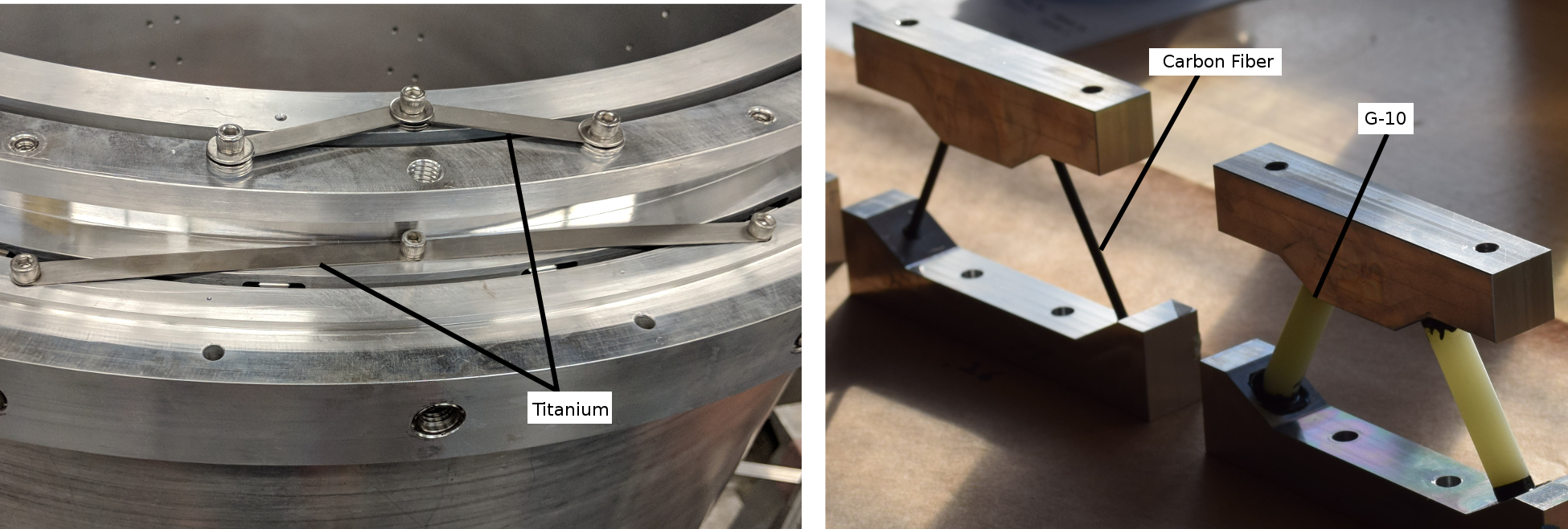}
	\caption{\textbf{Left:} The top end titanium supports. These thin strips
	are flexible along the axial direction of the cryostat while rigid enough
	to maintain concentricity of the stages. \textbf{Right:} Example carbon fiber (left) and G10 (right) supports. These
	supports sit at the back end of the cryostat and span the gap between the
	vacuum jacket to 50K stage (G10) and between the 50K and 4K stage (carbon
	fiber). Each rod is held in place with Stycast epoxy.}
	\label{fig:supports}
\end{figure}

The radiation absorbed by the interior stages is reduced by the use of
multi-layer insulation (MLI) wrapped around the outside of the 50K and 4K
stages. These MLI blankets are composed of many aluminized Mylar\textsuperscript{\textregistered} layers
separated by polyester spacer layers. Where there is insufficient room for
uncompressed insulation, we plan to use low emissivity aluminum tape on parallel
faces to decrease the radiation absorbed by the lower temperature surface.
A low thermal conductivity support system keeps the three stages thermally
insulated from each other and provides structural support. The front end of
each shell is constrained by thin Ti-Al-4V straps which allow flexing along the
axial direction of the cryostat to absorb differential thermal contraction. At
the back end, the 50K and 4K stages are each supported by six trusses. Each
truss has two high tensile strength, low thermal conductivity rods bonded to
aluminum blocks with Stycast\textsuperscript{\textregistered} epoxy. We use G10-FR4 for the back-end supports
between the vacuum shell and the 50K radiation shield but switch to carbon
fiber between the 50K and 4K shells due to the latter's lower thermal
conductivity at low temperatures. Figure~\ref{fig:supports} shows fabricated
examples of the front and back-end supports.

\begin{table}[b]
	\center
\begin{tabular}{|c||c|c|}
	\hline
	 & 50k & 4k \\
	 \hline
	Radiation & 2.75W & 0.007W \\
	Titanium Supports & 0.95W & 0.081W \\
	G10 Supports & 1.9W & \\
	Carbon Fiber Supports & & 0.076W\\
	Cryocables & 1.43W & 0.12W \\
	Infrared Loading & 14W & 0.154W \\
	\hline
	Total & 21.04W & 0.386W \\
	\hline
\end{tabular}

	\caption{Calculated loading from different components in the \biceparray\
	cryostat. Thermal conductivity numbers taken from the NIST cryogenics
	resource group\cite{nist} for most materials. Runyan and
	Jones\cite{runyan2008} provide thermal conductivity measurements for
	Graphlite carbon fiber at low temperatures. Loading due to cryocables
	assumes twenty 100-way MDM cables. This number will be higher for the
	150GHz receiver and lower for the 30/40 GHz receiver. Infrared loading
	refers to optical power absorbed by the optical components. These numbers
	are taken from the performance of \bicep3 and scaled if necessary to
	account for slight changes in active area.}
	\label{tab:loading}

\end{table}

In addition to providing radiation shielding and mount points for low
temperature optics, the 50K and 4K stages provide natural heat sinks for the
cryocables that connect the sub-Kelvin electronics to the exterior
room-temperature data acquisition system. By sinking the cryocables to the 50K
stage, the conductive loading beyond the 50K stage is significantly reduced.
Likewise, the heatsinks on the 4K stage further reduce the conductive
loading into the sub-Kelvin assembly.

\section{Copper Braid Heat Straps}

\begin{figure}[t]
\center
\includegraphics{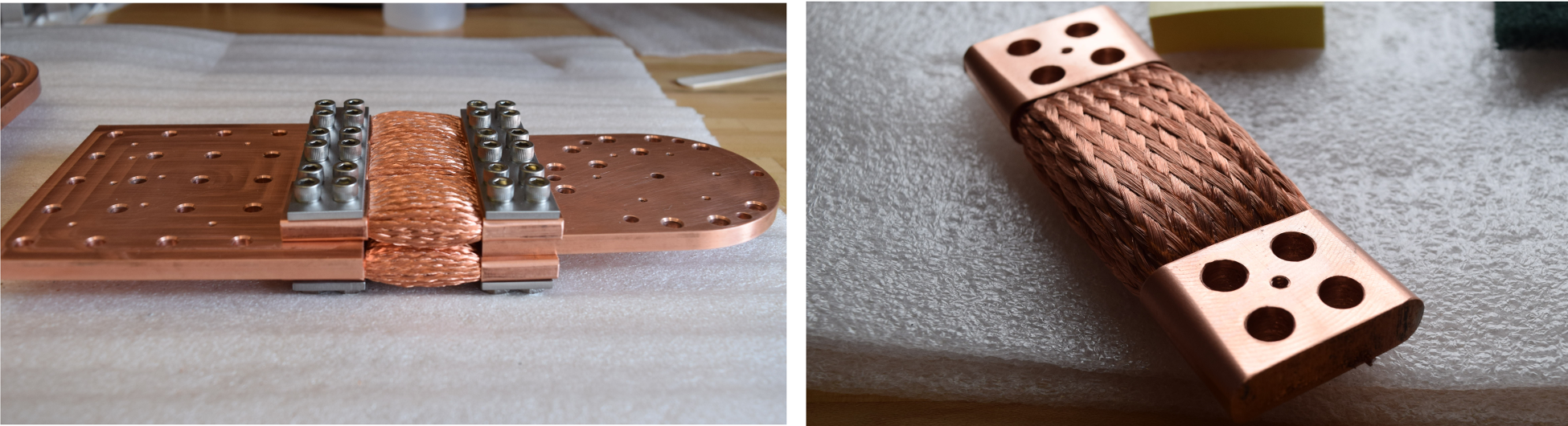}
\caption{\textbf{Left:} A completed heat strap composed of 6 braid assemblies. A
thin stainless steel plate distributes the force of the bolts and nuts across the surface
of the braid ends. \textbf{Right:} A single 7-layer braided strap. This strap
is created by inserting layers of OFHC braid into a short OFHC tube and
compressing under 20 tons of pressure in a hydraulic press.}
\label{fig:heatstrap}
\end{figure}

The heat straps connecting the pulse tube cooler to the 50K and 4K stages of
the cryostat need to have large thermal conductance but also be fairly
flexible to suppress vibrations transmitted to the focal plane. \biceparray\ 
plans to use custom made oxygen-free high thermal conductivity (OFHC) copper
assemblies each composed of multiple straps. As shown
in Fig.~\ref{fig:heatstrap} each heat strap consists of two end blocks
connected by a series of multi-layered braided wire straps. The braided straps
comprise seven layers of OFHC braid pressure fused into a small diameter OFHC
pipe section on either end. The pressure fusing is performed by a hydraulic
press under a load of 20 tons with lateral constraint provided by a steel
die. We have been able to achieve thermal conductance of $G=600
\frac{\text{mW}}{\text{K}}$ per strap at $4$K in laboratory tests.

The heat straps in the \biceparray\ cryostat combine a number of these straps
to achieve high total thermal conductance. Two layers of braided straps are
sandwiched around an OFHC plate on both ends. These plates provide mounting
interfaces to the rest of the cryostat and the pulse tube cooler. Stainless
steel plates on the top and bottom sides of this interface allow the use of
1/4'' stainless steel bolts to create a high pressure joint and reduce thermal
contact resistance. Figure~\ref{fig:heatstrap} shows a fully assembled heat
strap assembly that interfaces between the 4K radiation shield and the second
stage of the pulse tube.

\section{Mount and Operations}

The larger size of the \biceparray\ as compared to the \keckarray\ it replaces
requires a larger motorized platform for operation. The new \biceparray\ mount uses the same three-axis design as the previous
\bk\ experiments which augments the azimuth and elevation axes with
rotation about the boresight of the array. A
cross section of the new mount assembly is shown below in Fig.~\ref{fig:bamount}.
As with previous \bk\ experiments, the cryostats
are enclosed within an accordion-like environmental shield which co-rotates in
azimuth and flexes as the mount tips in elevation. A separate forward
plate provides a mounting interface for four absorbative optical baffles---one per
receiver---and co-rotates with the receivers about the array boresight.

\begin{figure} [hb]
	\begin{center}
		\includegraphics{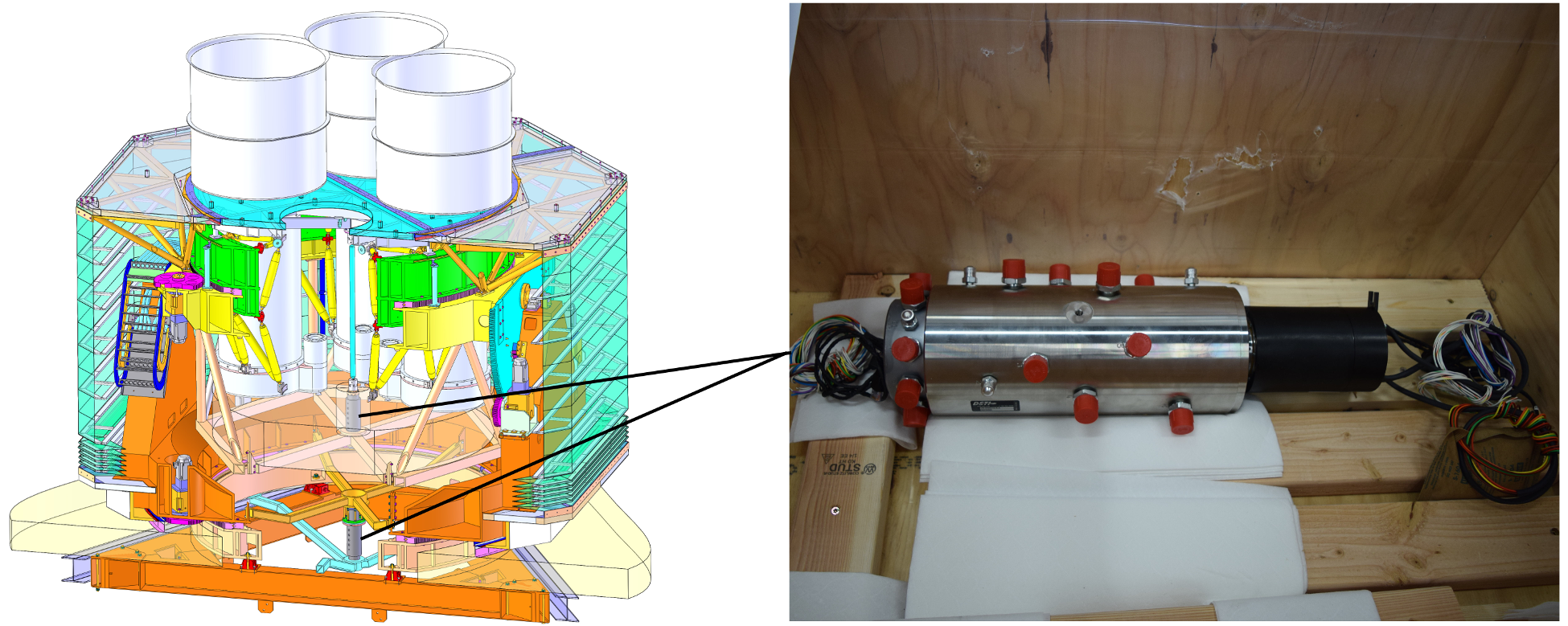}
	\end{center}
	\caption{\textbf{Right:} A CAD rendering of the new \biceparray\ mount and cryostats. The surrounding
	accordion-like environmental shield is shown in teal while the two rotary
	unions are depicted in gray and can be seen along the central axis of the
	mount. The large cylinders on the top side of the mount assembly are
	the co-moving absorptive optical baffles that reduce sidelobe response. A fixed
	ground shield (not shown) surrounds this mount assembly and reduces pickup from
	ground-fixed sources. Black lines indicate the two rotary unions.
	\textbf{Left:} One of the rotary unions from DSTI\cite{dsti}. Red caps denote the
	locations of helium lines The electrical and data connections
	can be seen exiting either end of the union.}
	\label{fig:bamount}
\end{figure}

The \biceparray\ mount includes two separate rotary unions which allow
continuous rotation about the azimuth axis and array boresight without the need for a
cable wrap (see Fig. \ref{fig:bamount}). These rotary unions were designed at DSTI\cite{dsti}, and each contain 10
helium channels. Eight of these connect the pulse tubes and their compressors,
while two channels serve as pressure guards. An additional nitrogen channel
provides a pressurized environment on front end of the cryostats which prevents water absorption into the window material.  The lip rings at the
ends of the unions additionally provide data and power connections to
electronics across separately rotating stages of the mount. These rotary unions
allow the helium compressors---required to operate the pulse tube coolers---to
sit well below the mount structure in the stationary tower. Helium lines route
upwards into the lower (ground fixed) half of the first rotary union and then
out through the upper half which rotates in azimuth along with the receivers.
The hoses from the upper half are then routed through a short cable chain that
provides flexure when rotating in elevation. The second rotary union is then
similarly connected between the elevation and boresight stages.

\biceparray\ will consist of four receivers observing in 6 frequency bands.
Two receivers will continue to observe in the 95 and 150 GHz bands where the
\bk\ maps are deepest and where combined foreground signal is at a minimum.
These will be augmented by two dual-band receivers at 30/40 GHz and 220/270
GHz.  The 30/40 GHz receiver will extend the observations into two new bands at
which the synchrotron foreground is expected to dominate. The \keckarray\ is
already observing in the 220 and 270 GHz bands.  However with significantly
increased throughput, and a detector count of over 8 times the entire
\keckarray, the dual band 220/270 GHz \biceparray\ receiver will rapidly
eclipse current sensitivity. In only a few days of observation, this receiver
will surpass the dust sensitivity of the \planck\ 353 GHz data in the \bk\
field. With the increased sensitivity at 95 and 150 GHz, these two additional
receivers will be required to push constraints on polarized emission from
galactic synchrotron and dust further than the currently available data.

\begin{figure}[b]
\center
\includegraphics[scale=0.5]{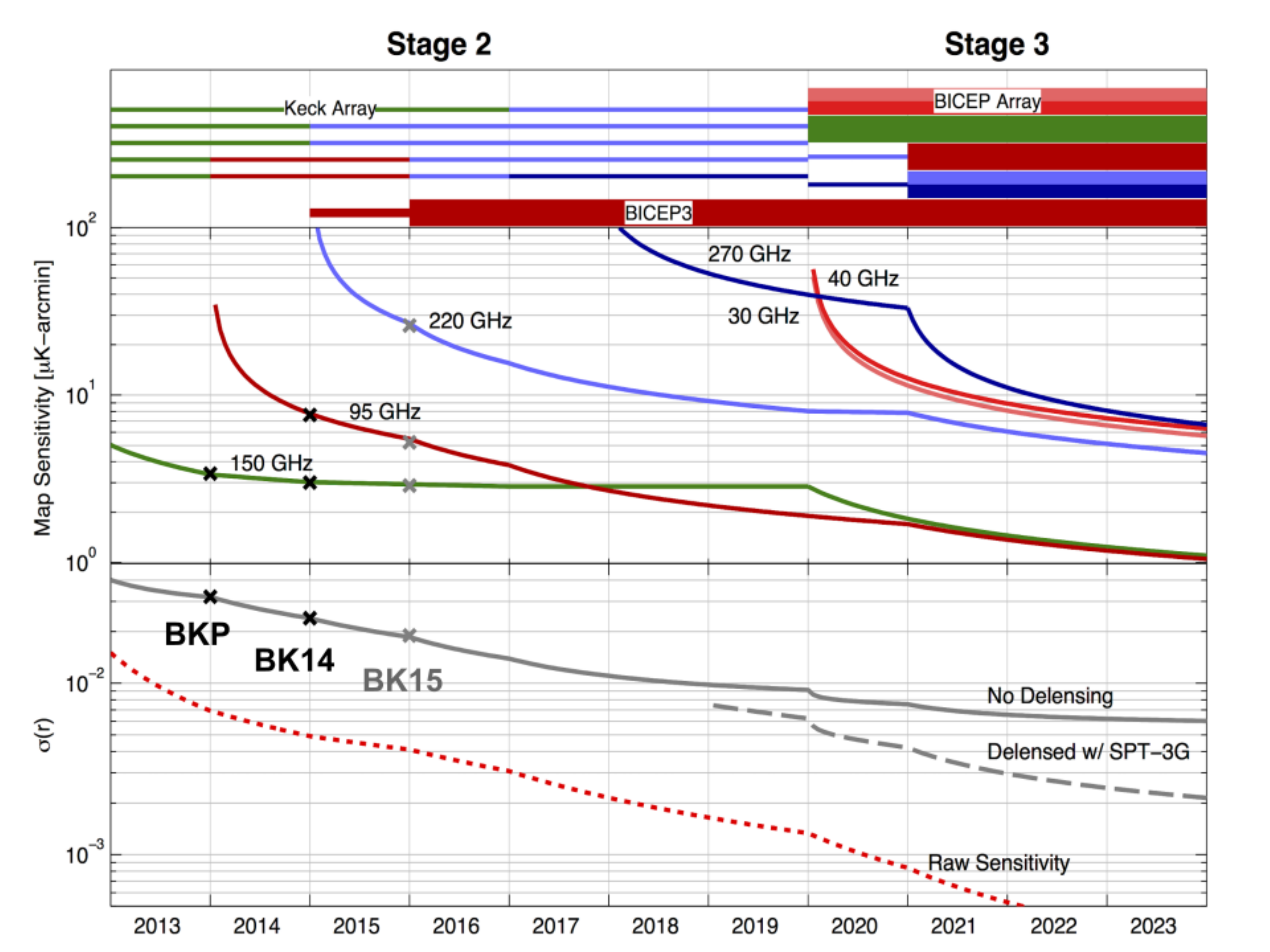}
\caption{Achieved and projected sensitivity for the current and planned \bk\
program. \textbf{Top:} A representation of receiver throughput at different
frequencies. \textbf{Middle:} Achieved and projected map depth at various
frequencies as a function of time. Black X's mark achieved sensitivities in the
BKP\cite{bkplanck} and BK14\cite{bk_vi} papers, while gray X's show achieved
sensitivity in the upcoming BK15 paper. \textbf{Bottom:} Sensitivity to $r$,
marginalizing over the seven parameter foreground model as described in BK14.
Raw sensitivity in the absence of foregrounds is also shown.}
\label{fig:projections}
\end{figure}

The observing power of \biceparray\ will be concentrated on the same $\sim400
\text{ deg}^2$ ($\sim 600 \text{deg}^2$ for \bicep3 generation) patch of sky as the existing \bk\ data. By directly
observing cosmological foregrounds with the new dual band receivers in the
patch at which the \bk\ data is already the deepest, we will be able to
directly constrain these foregrounds in our own patch of sky, significantly
reducing the effect of any spatial variation in the foregrounds' spectral
energy distribution. Increasing foreground constraints will be complemented by
simultaneously increasing sensitivity to $r$ with the single band receivers.
Figure~\ref{fig:projections} shows the sensitivity projections for \biceparray.
All projections are based on achieved sensitivity and as such, build in real
world inefficiencies such as detector yield, weather, and other factors which
decrease overall sensitivity. By the end of the program, \biceparray\ is
projected to constrain $r<0.008$ ($95\%$ confidence), $\sigma (r)\sim0.003$.

\section{Acknowledgements}

The \bk\ projects  have  been  made possible through a series of grants from the
National Science  Foundation  including  0742818,  0742592,  1044978, 1110087,
1145172, 1145143, 1145248, 1639040, 1638957, 1638978, 1638970, \& 1726917, by
the Gordon and Betty Moore Foundation, and by the Keck Foundation.  The
development of antenna-coupled detector technology was supported by the JPL
Research and Technology Development Fund and NASA Grants 06-ARPA206-0040, 10-
SAT10-0017, 12-SAT12-0031, 14-SAT14-0009 \& 16-SAT- 16-0002.   The  development
and  testing  of  focal  planes were supported by the Gordon and Betty Moore
Foundation at Caltech.  Readout electronics were supported by a Canada
Foundation for Innovation grant to UBC.  The computations in this paper were
run on the Odyssey cluster supported by the FAS Science Division Research
Computing Group at Harvard University.  The analysis effort at  Stanford  and
SLAC  is  partially  supported  by the  U.S.  DoE  Office of  Science.   We
thank  the  staff  of the U.S. Antarctic Program and in particular the South
Pole Station without whose help this research would not have been possible.
Most special thanks go to our heroic winter-overs  Robert Schwarz  and  Steffen
Richter.   We thank all those who have contributed past efforts to the
\bk\ series of experiments, including the \bicep1 team.

\clearpage

\bibliography{cmb,misc,bicepkeck}
\bibliographystyle{spiebib}

\end{document}

%% file: defs.tex
\def\bicep{\textsc{Bicep}}

\def\biceparray{{\sc Bicep} Array}

\def\keck{{\it Keck}}
\def\keckarray{{\it Keck Array}}
\def\bk{\bicep/\keck}

\def\planck{{\it Planck}} 

